\documentclass[12pt,preprint]{aastex}
\shorttitle{Nonlinear Instability of kink oscillations due to shear motions}
\shortauthors{Terradas et al.}

\begin{document}

\title{Nonlinear Instability of kink oscillations due to shear motions}

\author{J. Terradas, J. Andries\altaffilmark{1}, M. Goossens}

\email{jaume@wis.kuleuven.be}

\affil{Centre for Plasma Astrophysics and Leuven Mathematical Modeling and
Computational Science Centre \\ Katholieke Universiteit Leuven, Celestijnenlaan
200B, B-3001 Leuven, Belgium
}

\and

\author{I. Arregui, R. Oliver, J. L. Ballester}

\affil{Departament de F\'\i sica, Universitat de les Illes Balears, 
E-07122 Palma de Mallorca, Spain}


\altaffiltext{1}{Postdoctoral Fellow of the National Fund for Scientific
Research--Flanders (Belgium) (F.W.O.-Vlaanderen).}

\begin{abstract}

First results from a high-resolution three-dimensional nonlinear numerical study
of the kink oscillation are presented. We show in detail the development of a
shear instability in an untwisted  line-tied magnetic flux tube. The instability
produces significant deformations of the tube boundary. An extended transition
layer may naturally evolve as a result of the shear instability at a sharp
transition between the flux tube and the external medium. We also discuss the
possible effects of the instability on the process of resonant absorption when
an inhomogeneous layer is included in the model. One of the implications of
these results is that the azimuthal component of the magnetic field of a stable
flux tube in the solar corona, needed to prevent the shear instability, is
probably constrained to be in a very specific range.

\end{abstract}


\section{Introduction}

Coronal loops and filament threads are magnetic flux tubes whose field lines are
anchored to the dense photosphere. Usually these structures are modeled as
circular tubes whose properties change mainly in the radial direction with a
sharp or a smooth transition. Using the ideal magnetohydrodynamic (MHD)
equations the eigenmodes of cylindrical magnetic tubes can be calculated. A
fundamental eigenmode of oscillation is the kink mode, with an azimuthal number
$m=1$, which produces a transverse displacement of the tube.  Examples of the
calculations of the kink mode for sharp interfaces can be found in
\cite{spruit81,edrob83,cally86} and for smooth transition layers in 
\cite{gooss92, rudrob02, vand04, terr06}. In the last case, the coupling of
compressional fast magnetohydrodynamic (MHD) waves and shear Alfv\'en waves
leads to the formation of resonances in the inhomogeneous layers and this
mechanism is a possible candidate to explain the damping of transverse coronal
loop oscillations \citep{hollyang88, rudrob02, gooss02} and filament threads
\citep{arr08}. 

So far, most of the theoretical studies about the kink mode are in the linear
regime and second order perturbations are neglected in the MHD equations,
however, there are some observational indications suggesting that this
assumption might be not fully justified in all the oscillating loops
\citep[][]{terrof04}. Nonlinearity adds new and interesting effects. For
example, the ponderomotive force creates a flow along the field lines
\citep{rank94,tik95}, and for the standing kink oscillation it tends to
accumulate mass at the loop apex \citep[][]{terrof04}. When gas pressure is
taken into account, the otherwise unlimited accumulation of mass is prevented
by the pressure forces which limit the secular growth and produce saturation in
the density. 

Another possible consequence of the nonlinearity is the generation of
instabilities. It is well known that even in the linear regime, a sharp
interface between two media in relative motion is liable to the Kelvin-Helmholtz
instability (KHI). In the presence of a magnetic field, KHI of a flow parallel
to the field lines in a homogeneous medium sets in if the velocity jump exceeds
the maximum Alfv\'en speed. The criteria for the KHI for an axial flow is
modified in the presence of a boundary layer of finite width, which also induces
the possibility of resonant flow instabilities
\citep[see][]{holl90,yangh91,and01a,and01b}. The effect of an azimuthal flow is
less explored in the literature. \citet{heyvpri83,browpri} showed that azimuthal
shear motions in the presence of a smooth transition layer can be KH unstable.
The most likely place for the KHI to occur is where the velocity is largest and
the magnetic field is perpendicular to the velocity \citep[see for
example][]{rank93}. For the fundamental standing kink mode this is precisely the
antinode of the velocity, located at half the loop length from the photosphere. 

In this Letter we report on the nonlinear evolution of the kink oscillation of a
tube without twist. The full nonlinear three-dimensional ideal MHD equations are
solved numerically. From the simulations we investigate the motions in the
magnetic tube and show the first results of the development of the nonlinear
shear instability at the tube boundary.

\section{Tube Model and Initial Conditions}\label{model}

The system under consideration is a compressible plasma that obeys the general
equations of MHD. The equilibrium magnetic field is straight, uniform, and
pointing in the $z-$direction. In a cylindrical coordinate system with the axis
along the magnetic field, the density of the circular tube changes with the
radial coordinate from $\rho_{\rm in}$ inside the tube to $\rho_{\rm ex}$ in the
coronal medium through an inhomogeneous layer of width $l$. The length of the
loop is $L$ and the mean radius $R$. To have magneto-hydrostatic equilibrium the
gas pressure is uniform, and the plasma$-\beta$ is chosen to be small, $5\times
10^{-2}$ (we assume a sound speed of $c_{\rm s}=0.2\,v_{\rm Ai}$, where $v_{\rm
Ai}$ is the Alfv\'en velocity inside the tube). In this model the role played by
the non-zero gas pressure is important since it avoids the continuous
accumulation of mass at the tube half length due to the nonlinear ponderomotive
force. We apply line-tying conditions at the planes
located at $z=0$ and $z=L$ to mimic the anchoring in the photosphere and to
obtain a standing mode.

The system is perturbed with the spatial structure of the linear fundamental
standing kink eigenmode ($m=1$, and longitudinal wavenumber $k_z=\pi/L$) of a
magnetic tube with a discontinuous density profile. This allows exciting a
single eigenmode (in the linear regime) and avoiding a significant excitation of
leaky modes. For this disturbance the radial component of the velocity is
continuous across the loop boundary while the azimuthal component has a jump at
the boundary. The initial perturbation is the seed of the instability in the
nonlinear regime. The amplitude of the initial velocity perturbation  at the
center of the tube ($x=0$, $y=0$) is $v_0$. Another relevant magnitude in our
analysis is the axial component of the vorticity, $\Omega_z=(\nabla\times {\bf
{v}})\cdot {\bf {\hat e}_z}$. The initial kink perturbation is a vortex sheet at
the tube boundary (see Fig.~\ref{density}a).

The radial, azimuthal, and longitudinal dependence of the initial perturbation
are transformed to the 3D Cartesian system of our computational box. Given the
initial condition the time-dependent nonlinear MHD equations are
numerically solved. We have used an explicit high-order numerical scheme
($4th-$order in time and $3th-$order in space) based on the method of lines to
solve the equations in conservative form \citep[see][for further
details]{terr08}. Due to the resolution requirements the parallelized version of
the code has been run in a cluster of machines. We have used a grid resolution
of $512 \times 512 \times 100$ points since the small scales are in the $x-y$
plane while the solution is smooth in the $z-$direction. 

\section{Tube Evolution}\label{tube}

We have started the analysis of the evolution of the tube with very small
amplitudes of the initial perturbation ($v_0\ll v_{\rm Ai}$). From the fully 3D
problem we recover the linear results, i.e. the loop oscillates around the
equilibrium position as a whole with the kink mode frequency. If instead of a
sharp density transition between the tube and the external medium we use a
smooth density profile (we have used thick layers to have enough grids points
inside the resonant layer), then the tube attenuates due to the energy
conversion between the fast MHD waves and the Alfv\'en waves in the layer. We
have found that the period and damping time are in good agreement with the
theoretical linear predictions.

The behavior of the system changes completely when we are in the nonlinear
regime. The first effect is the generation of flows along the tube axis produced
by the nonlinear terms. However, these flows  are much smaller (in the regime
considered here) than the local Alfv\'en speed and therefore unable to generate
an instability. The dynamics of the system is dominated by the azimuthal flows
at the tube boundary. 

\subsection{Sharp transition layer} 

We first consider a sharp transition between the tube and the external medium.
The evolution of the density and the longitudinal component of the vorticity of
a representative case in a weak nonlinear regime is shown in
Figure~\ref{density} at different time intervals. To visualize the results of
the 3D simulations we concentrate on the plane at half the tube length where we
expect the strongest nonlinear effects. The initial perturbation  produces a
lateral displacement of the tube in the $x-$direction. As in the linear regime
the tube starts to oscillate around the initial position, but small length
scales quickly develop at the boundary. We can appreciate this, for example,
around the points $x=0$ and $y=\pm R$ (see Fig.~\ref{density}b), which is the
position where $v_x$ has a maximum jump. These small scale structures grow with
time (see Fig.~\ref{density}c) and several rolls form at the loop boundary. Note
that around $x=\pm R$ and $y=0$ the density is almost undisturbed because there are
no shear motions at this position. At later stages of the evolution
(Fig.~\ref{density}d) the small spatial scales are still localized at the
boundary and eventually the system reaches a saturation state. As a result of
the instability the overall shape of the tube at the boundary has been
considerably altered and a rather inhomogeneous layer has been generated. The
changes at the tube boundary are also clear in the vorticity. In the early phase
the initial vortex sheet, located at the boundary, shows small and localized
deformations (Fig.~\ref{density}b). Since the tube is oscillating the flow
changes sign in each oscillation,  and the vortex sheet evolves in a complex
way, showing a very undulated shape (Figs.~\ref{density}c and \ref{density}d).

\subsection{Smooth transition layer}

Now we assume that the equilibrium already has a transition region. The main
difference with respect to the previous case is that the resonant absorption
process induces shear motions at the inhomogeneous layer due to phase mixing.
The results are represented in Figure~\ref{density1} for two different widths
of the layer at a given time instant (same as in  Fig.~\ref{density}c). We see
that the instability is also present (Fig.~\ref{density1}a), but now it is less
developed in comparison with the sharp transition case (Fig.~\ref{density}c).
We also see (Fig.~\ref{density1}b) that  the thicker the layer the slower the
growth-rate of the instability (in Fig.~\ref{density1}b the instability is
still not present). Since the boundary of the tube eventually changes its
shape due to the instability, a question that arises is how this affects the
process of resonant absorption. We have calculated the damping rate (damping
time over period) of the tube  from the simulations for the two cases
considered in Figure~\ref{density1} (calculating the displacement of the
central point of the tube). The numerical estimates give values around 1.9 for
$l=0.5R$ and 1.2 for $l=R$, while the linear values of the damping rates based
on eigenmode calculations are 2.4 and 1.1, respectively. The differences are
small,  indicating that the instability, for the particular parameters
considered here, does not change much the efficiency of the energy conversion. 

\section{Discussion and Conclusions}\label{concl}

The numerical results shown in this Letter indicate that the shear motions
involved in the kink oscillations of a magnetic tube might be unstable. The
instabilities are found to create small length scales in the azimuthal direction
and grow rapidly in time. For a cylindrical discontinuous interface between two
homogeneous stationary rotating fluids the following growth rates (imaginary
part of the frequency) of the Kelvin
Helmholtz instability can be readily derived: \begin{equation}\label{kh}
\omega_i^2=\frac{\rho_{\rm in}\rho_{\rm ex}}{(\rho_{\rm in}+\rho_{\rm
ex})^2}\frac{m^2}{R^2} 4   v_0^2-2 k_z^2 \frac{B_0^2}{\mu (\rho_{\rm
in}+\rho_{\rm ex})}. \end{equation} Here $2v_0$ is the amplitude of the velocity
shear at the boundary. The derivation involves the assumption that the azimuthal
length scales are much smaller than the longitudinal ones (which is the case for
the instabilities that formed in our study). Clearly the background equilibrium
for which  equation~(\ref{kh}) is obtained is very different from the shear
motions associated to the kink mode oscillations, which depend on time, $\phi$
and $z$. Nevertheless, the small scales and the localization of the
instabilities in the azimuthal direction, and the fast growth rates, suggest
that equation~(\ref{kh}) may be considered as a local analysis and we need not
to worry about the azimuthal and time dependence of the shear motions. For the
parameters considered in \S\ref{tube}, the first unstable mode corresponds to
$m=5$ (which satisfies the assumption of azimuthal localization), and its growth
rate is $1/\omega_i\approx 5.0\tau_{\rm A}$. The important point here is that it
is smaller than the kink period, around $16.3\tau_{\rm A}$ (satisfying the
assumption of localization in time). Moreover, this calculation also clarifies
that the most unstable modes will have small azimuthal length scales but large
longitudinal length scales, since this minimizes the stabilizing force induced
by the bending of the field lines. However, the large longitudinal length scales
prevent to interpret equation~(\ref{kh}) as a local approximation in $z$ since
the velocity shear is less than $2v_0$ throughout most of the loop. Growth rates
are thus expected to be somewhat smaller, and instability will set in only for
azimuthal length scales smaller than those predicted by equation~(\ref{kh}). An
appropriate analysis thus needs to take into account the longitudinal variation
of the shear profile and hence necessarily involves a 2D model. 

\begin{figure*}[!ht]
\center{
\includegraphics[width=6.4cm]{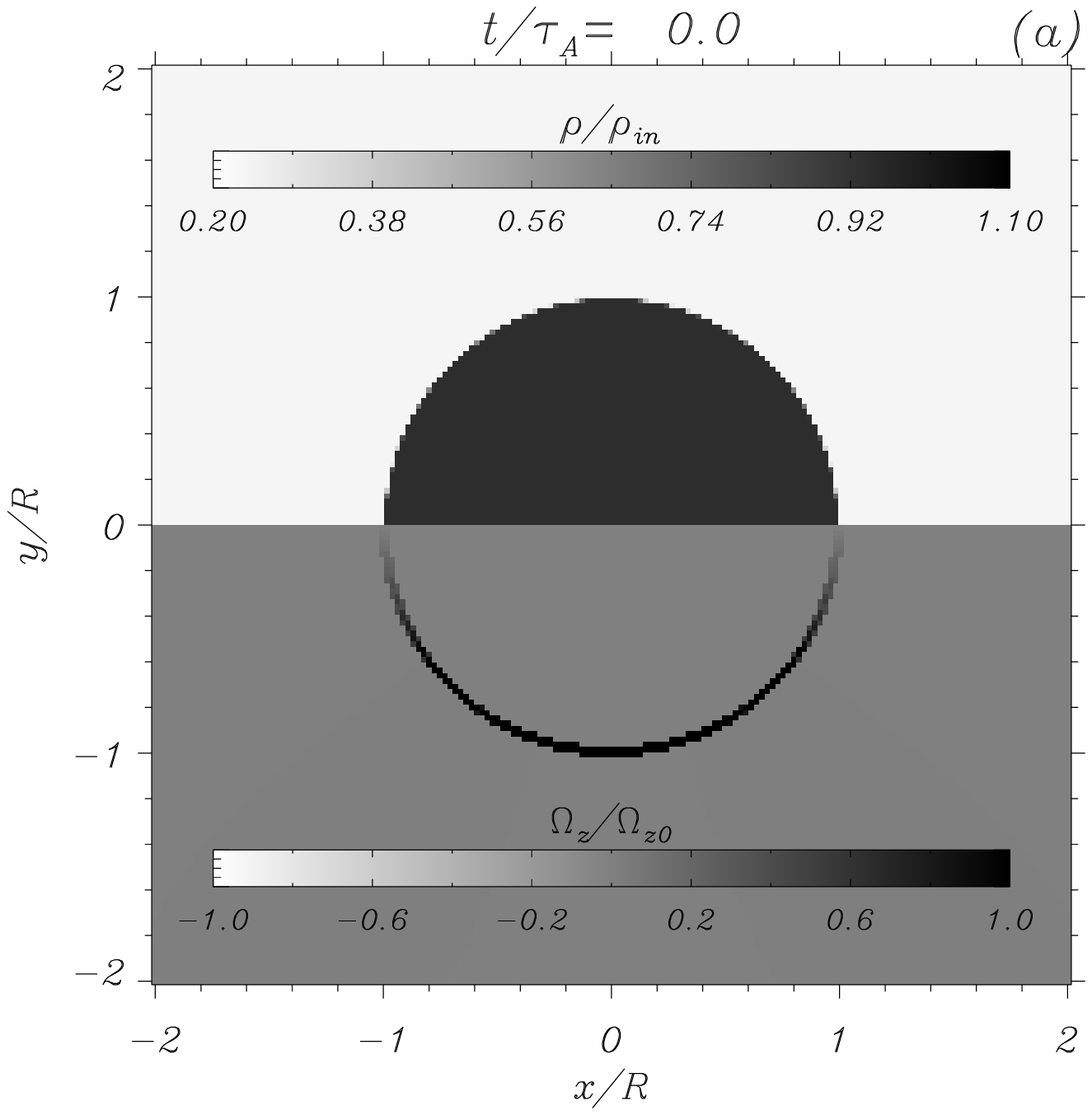}
\hspace{1cm}
\includegraphics[width=6.4cm]{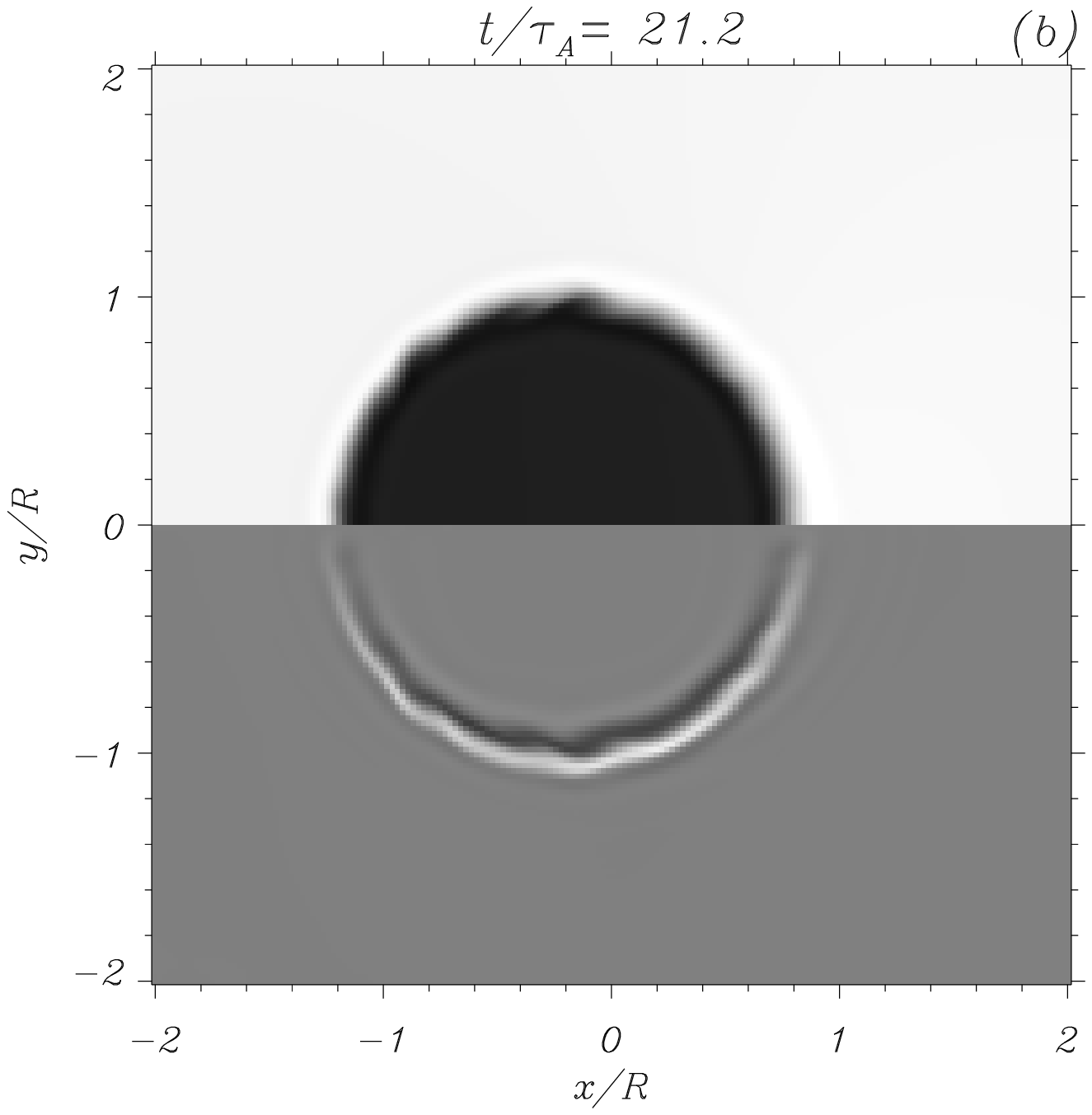}
\\
\includegraphics[width=6.4cm]{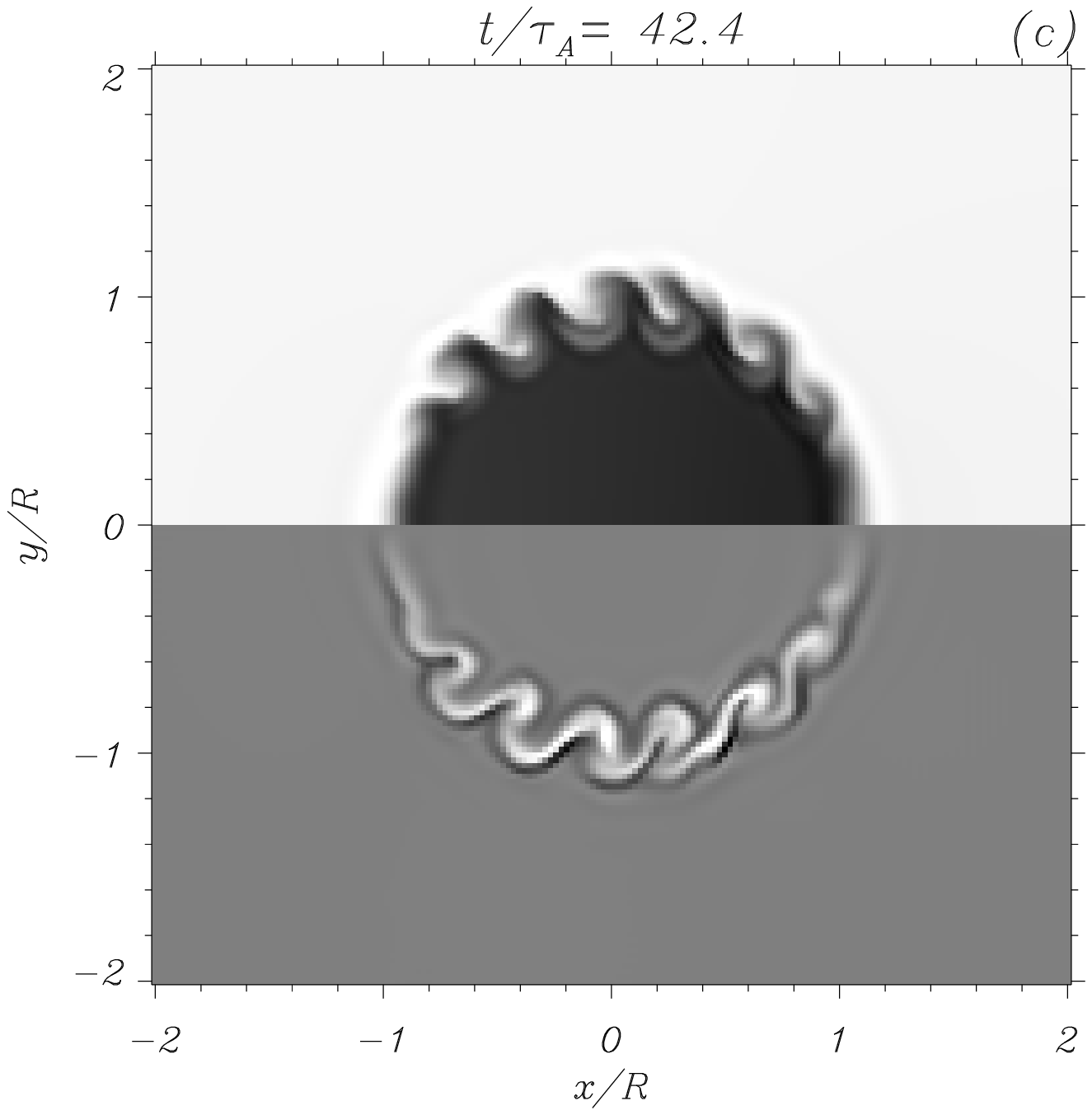}
\hspace{1cm}
\includegraphics[width=6.4cm]{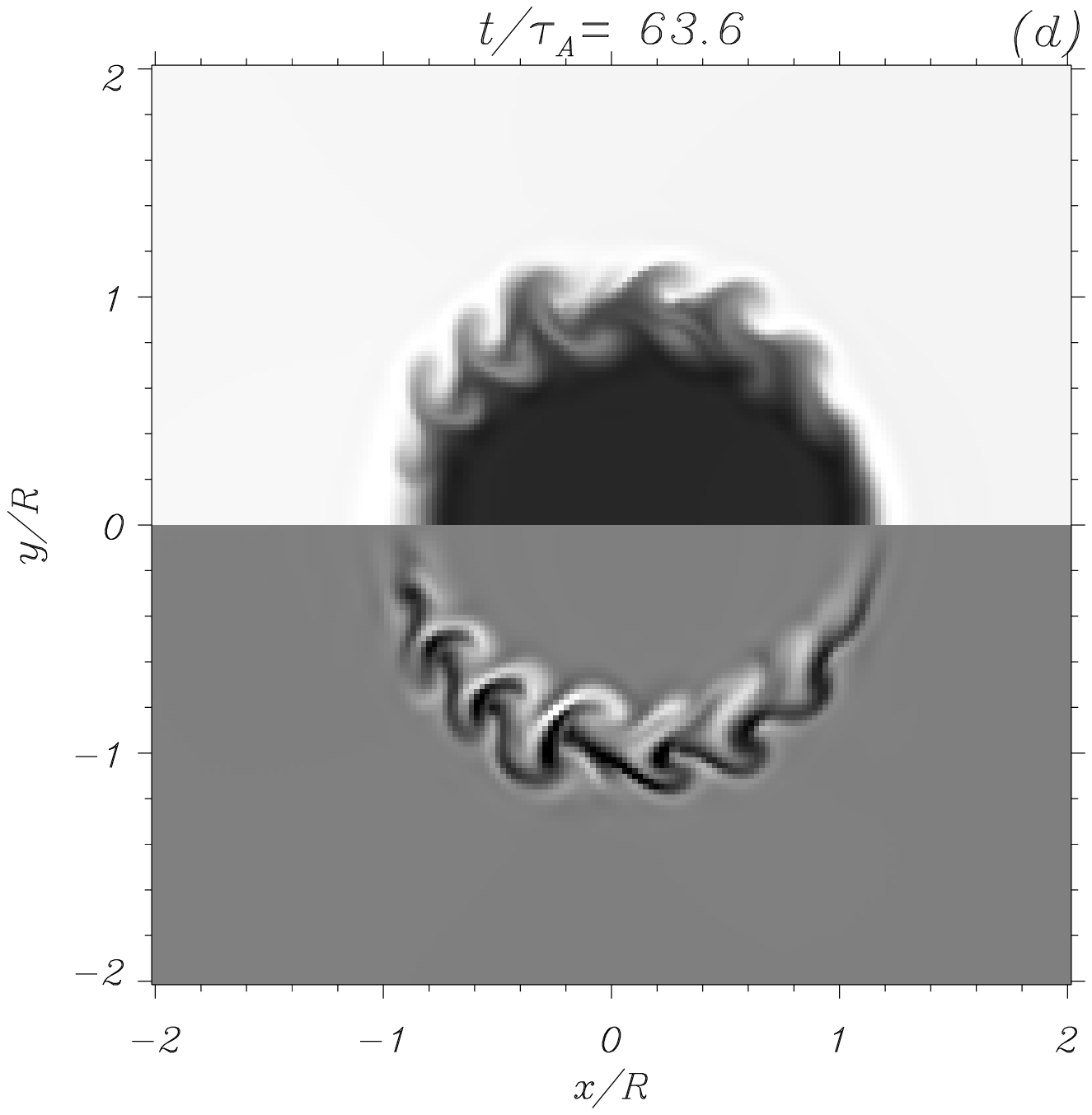}
}
\caption{ 
\small  The top half of each panel displays the density (symmetric respect to
$y=0$) and the bottom half the vorticity (antisymmetric respect to $y=0$) at
$z=L/2$ for
different times. In this simulation the following parameters have been used: 
$L=10R$, $\rho_{\rm in}/\rho_{\rm ex}$=3, $v_0=0.1v_{\rm Ai}$, and a full domain of
$[-6R,-6R]\times[-6R,6R]\times[0,10R]$. Lengths are normalized to the loop
radius, $R$ (typically of the order of $4\,000$\,km), velocities to the internal Alfv\'en velocity, $v_{\rm
Ai}=B_0/\sqrt{\mu \rho_{\rm in}}$ (of the order of $1\,000\,{\rm km\,s^{-1}}$), and time to the Alfv\'en crossing time,
$\tau_{\rm A}=R/v_{\rm Ai}$. {\em This figure is available as an mpeg animation
in the electronic edition of the Journal.}}  \label{density} \end{figure*}

\begin{figure*}[!ht]
\center{
\includegraphics[width=6.4cm]{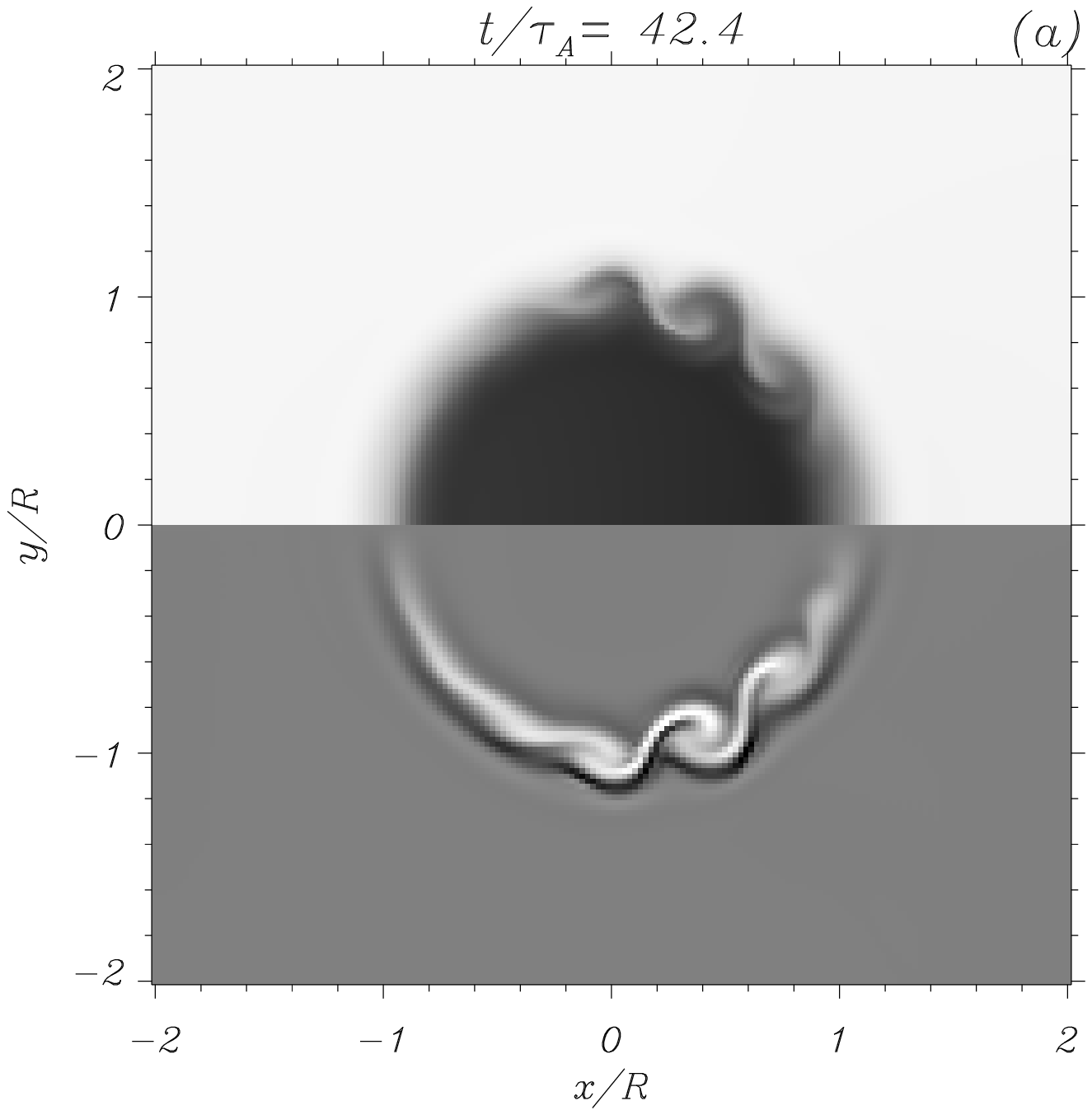}
\hspace{1cm}
\includegraphics[width=6.4cm]{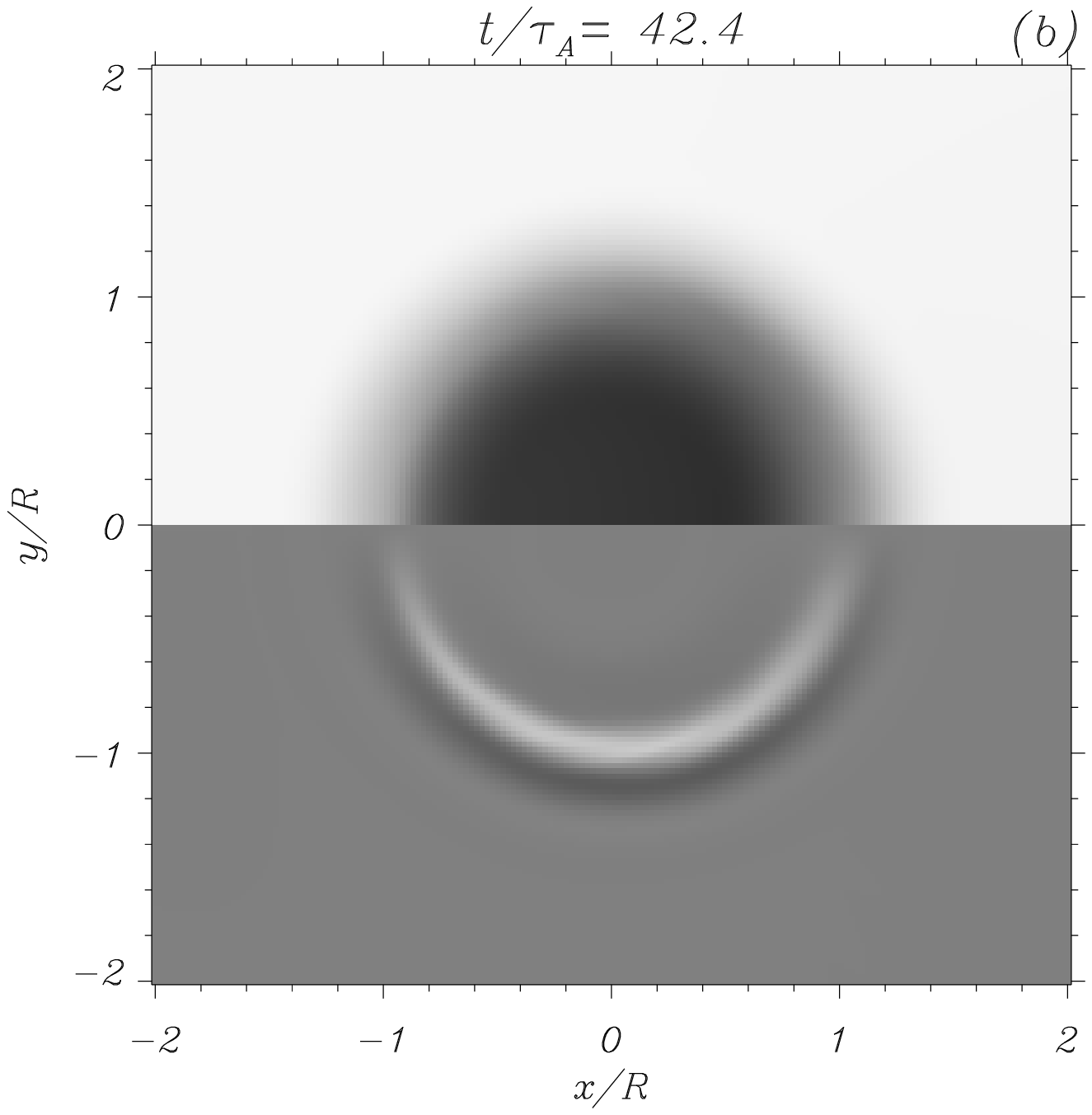}
}
\caption{ \small  Density and
vorticity (for the same time instant as in Fig.~\ref{density}c) for two
different widths of the inhomogeneous layer connecting the tube and the external
medium, {\bf
a)}
$l=0.5R$ and {\bf b)} $l=R$. All other parameters are the same as in
Figure~\ref{density}.}  \label{density1}
\end{figure*}

We have found that the evolution of the tube is very sensitive to the amplitude
of the initial perturbation. As expected, the larger the amplitude of the
initial perturbation the faster the development of the instability. By changing
the initial amplitude of the perturbation in a broad range, an extended
transition layer (even much larger than the developed in Fig.~\ref{density}) may
naturally evolve as the result of the shear instability of a sharp transition
between the flux tube and the external medium. In addition, for very large
amplitudes the tube might show severe deformations, a wake can even form behind
the tube and it can interact with the main body in each oscillation, the final
shape of the tube being rather irregular. 

When an inhomogeneous layer between the tube and the environment is included
then the motions are characterized by phase-mixed scales and the instability is
also present. Nevertheless, the instability develops more slowly than in the
sharp transition case. This is probably due to the fact that the thicker the
layer the later the generation of small length scales due to the phase mixing
process, and thus the onset of the instability. This situation seems to be
related to the development of the KHI for torsional Alfv\'en waves ($m=0$)
described by \citet{browpri} \citep[see also][]{walker}. Interestingly, for the
regime studied here (basically thick layers) the attenuation of the central part
of the tube due to resonant absorption is not significantly altered by the
changes at the boundary due to the shear instability. 

In the context of coronal loops an immediate question that arises from the
results presented here is why, up to now, there is no clear evidence of such
instability from the observation of oscillating loops. There are several partial
answers. Magnetic twist, not included in our model, might decrease or even
suppress the instability since the presence of a magnetic field component along
the flow stabilizes the KHI. Several examples of the stabilizing effect of a
helical magnetic field component can be found in rising tubes in the convection
zone \citep[see for example][]{fer96} or in stellar jets. On the other hand, a
tube with very large azimuthal magnetic field  is subject to the instabilities
of the linear pinch (typically for a twist larger than $2.5\pi$, although it
depends on the details of the equilibrium and boundary conditions). Therefore,
the azimuthal component of the magnetic field of a stable flux tube in the solar
corona is probably constrained to be in a specific range (being $2.5\pi$ an
upper bound for the twist).  Other factors that could explain the absence of
observational evidences of the instability are that it is not spatially resolved
with the current spatial resolution of the telescopes or simply that the
amplitude of the oscillating loops is not strong enough to develop the
instability. However, the tube displacement produced by the initial perturbation
in our simulations is of the order of and even smaller than the observed
amplitudes of oscillation in loops, so this last possibility seems to be ruled
out.

We have given a qualitative description of the shear instability, but a
quantitative analysis (with improved numerical resolution) about the growth
rates of the modes, and a detailed study of the effect of the instability on the
damping rates under different regimes is still needed. In addition, the
equilibrium configuration should be improved in several aspects. For example, a
more accurate model should incorporate a realistic variation of the density and
temperature from the photosphere to the corona. However, the inclusion of a
twisted magnetic field for the reasons mentioned before seems to be the most
relevant aspect that needs to be addressed.

\acknowledgements  J. Terradas is grateful to the Research Council fellowship
F/06/65 of the KUL. Spanish funding provided under grants AYA2006-07637 and
PCTIB-2005GC3-03 is acknowledged. We are grateful to Fernando Moreno-Insertis
for his comments and suggestions.

\clearpage

\end{document}